# Machines hear better when they have ears

**Deepak Baby**[a,1] **and Sarah Verhulst**[a]

[a]Department of Information Technology, Gent University, Technologiepark 15, 9052 Zwijnaarde, Belgium

**Deep-neural-network (DNN) based noise suppression systems yield significant improvements over conventional approaches such as spectral subtraction and non-negative matrix factorization, but do not generalize well to noise conditions they were not trained for. In comparison to DNNs, humans show remarkable noise suppression capabilities that yield successful speech intelligibility under various adverse listening conditions and negative signal-to-noise ratios (SNRs). Motivated by the excellent human performance, this paper explores whether numerical models that simulate human cochlear signal processing can be combined with DNNs to improve the robustness of DNN based noise suppression systems. Five cochlear models were coupled to fully-connected and recurrent NN-based noise suppression systems and were trained and evaluated for a variety of noise conditions using objective metrics: perceptual speech quality (PESQ), segmental SNR and cepstral distance. The simulations show that biophysically-inspired cochlear models improve the generalizability of DNN-based noise suppression systems for unseen noise and negative SNRs. This approach thus leads to robust noise suppression systems that are less sensitive to the noise type and noise level. Because cochlear models capture the intrinsic nonlinearities and dynamics of peripheral auditory processing, it is shown here that accounting for their deterministic signal processing improves machine hearing and avoids overtraining of multi-layer DNNs. We hence conclude that machines hear better when realistic cochlear models are used at the input of DNNs.**

neural networks | noise suppression | cochlear models | speech in noise

Speech recordings from realistic environments often have reduced intelligibility due to added degradations such as background noise and reverberation. Such degradations also adversely affect applications that rely on these speech recordings such as automatic speech recognition, speaker identification systems and hearing aids. It is often required to suppress these artifacts and *enhance* (or noise suppress) the speech recording before they are fed to such applications to yield a better performance and/or intelligibility (1–4).

Approaches to recover the speech signal from a noisy recording can be broadly classified as unsupervised and supervised techniques. Unsupervised techniques such as spectral subtraction (5) and Kalman filtering (6) make assumptions on the statistical properties of speech and noise that are often invalid on realistic recordings resulting in artifacts such as musical noise (7). Differently, supervised techniques learn the speech and noise statistics from a training dataset containing noisy speech and the underlying clean speech signal. With advances in machine learning, supervised approaches such as non-negative matrix factorization (NMF) (2) and code-book based systems (8) have shown great potential in improving the speech enhancement quality. This paper concentrates on supervised single-channel speech enhancement systems that aim to recover the clean speech signal from a noisy speech recording made with a single microphone.

Recent advances in supervised approaches which use deep neural networks (DNNs) have shown significant performance improvements over the existing techniques on a variety of complex machine learning tasks (9) including speech enhancement (10, 11). DNNs are comprised of multiple hidden layers which enable them to learn representations of data with multiple levels of abstraction (12). However, the performance of DNN-based systems are still far from that of humans especially for noises that are not present in the training data (dubbed *unseen* or *mismatched* noise conditions) and for negative signal-to-noise ratio (SNR) conditions. Although several core concepts of DNNs stem from the cortical processing in the human brain (13), most DNN-based speech enhancement systems still make use of engineered representations of the acoustic speech signal such as short-time Fourier transform (STFT) or Mel-integrated magnitude STFT (*FBANK* features) (11, 14–17). Since humans perceive speech remarkably well under a large variety of adverse listening conditions (18), we argue that using biophysically-inspired representations of speech might lead to a more generalizable DNN system.

This paper investigates whether biophysically-inspired speech representations can mitigate the poorer generalization capabilities of state-of-the-art DNN based speech enhancement systems employing fully-connected and recurrent neural networks (RNNs). To this end, we filter the acoustic signal using five different cochlear models and the resulting filtered representation of noisy speech are used as input features to the DNN. The considered cochlear models range from basic representations of cochlear filtering (e.g., Gammatone (GT) filter-bank energies (19)) to advanced biophysically-inspired non-linear models ( e.g., transmission line (TL) model (20)).

**Significance Statement**

Deep-neural-network (DNN) based techniques are arguably the most widely used tool in data science. However, for speech related applications such as automatic speech recognition and noise suppression, their performance is still far from that of humans especially under adverse noisy listening conditions. Most DNN systems are trained using engineered representations of speech that are different from how humans process it. This work investigates biophysically-inspired speech representations for DNN-based noise suppression systems that enhance the speech in a noisy speech recording. We show that such representations can improve the performance of the state-of-the-art DNN-based noise suppression systems especially for unseen noise conditions, suggesting that such representations result in machine hearing applications that are less sensitive to the noise type and level.

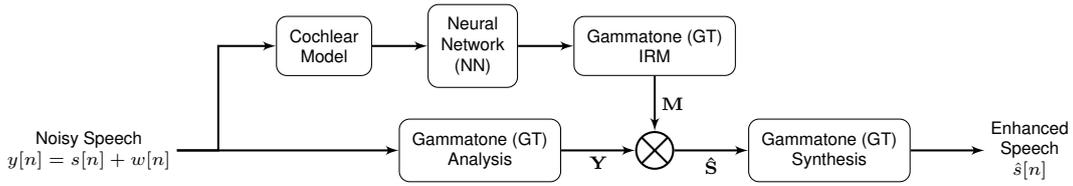

**Fig. 1.** Block diagram overview of a neural network-based speech enhancement system.

We evaluate the different DNN-based noise suppression systems for different noise types such as babble and factory noise under varying SNR levels ranging from $-3$ to $9$ dB and use objective speech enhancement metrics such as perceptual evaluation of speech quality (PESQ), segmental SNR (segSNR) and cepstral distance (CD) to quantify performance. We train and evaluate several state-of-the-art fully-connected and RNN-based speech enhancement systems, which receive the output of one of the five considered cochlear models as their input. We hypothesize that cochlear models which accurately describe the noise-suppression characteristics of human cochlear processing should yield better and more generalizable DNN-based noise suppression systems.

**Speech enhancement using NN**

We describe here how enhanced speech is obtained from the noisy speech recording in a single-channel speech enhancement system (see block diagram in Fig. 1). Let $s[n]$ and $w[n]$ be the time-domain clean speech and noise signals respectively, where $n$ is the sampling index. The goal of single-channel speech enhancement systems is to recover $s[n]$ from the noisy speech $y[n] = s[n] + w[n]$.

Since the frequency content in speech changes over time, a time-frequency (T-F) representation of speech over small, overlapping windows or *frames* is typically used. Let $\mathbf{Y}$, $\mathbf{S}$ and $\mathbf{W}$ be the magnitude T-F representations (neglecting the phase information) of $y[n]$, $s[n]$ and $w[n]$, respectively. The goal is thus to recover $\mathbf{S}$ from $\mathbf{Y}$. To obtain the time-domain enhanced speech $\hat{s}[n]$, an inverse T-F analysis is applied to the estimated $\hat{\mathbf{S}}$. Therefore, it is important to use a T-F representation that is invertible to obtain the time-domain speech signal. This work makes use of the popular Gammatone (GT) filterbank (comprised of $B$ channels) (19), followed by computing the envelope energies over frames as the T-F/spectrogram representation. Thus the goal is to enhance the noisy GT spectrogram using DNN-based approaches.

The bottom row of Fig. 1, depicts the GT analysis and synthesis pipeline. The clean speech GT representation is estimated by applying a *mask* $\mathbf{M}$ to the GT representation of noisy speech. This mask provides a scaling factor for every T-F point that extracts the GT representation of clean speech from that of the noisy speech. i.e., $\hat{\mathbf{S}}(b,f) = \mathbf{Y}(b,f) \cdot \mathbf{M}(b,f)$, where $\hat{\mathbf{S}}$ is the estimated clean speech GT representation. $b$ and $f$ denote the filter and frame indices of the T-F representation, respectively. The goal of the top row in Fig. 1 is thus to estimate this mask $\mathbf{M}$ from the noisy speech signal. We make use of DNNs that can predict the ideal ratio-mask (IRM) defined as

$$\mathbf{M}(b,f) = \frac{\mathbf{S}(b,f)}{\mathbf{S}(b,f) + \mathbf{W}(b,f)}. \quad [1]$$

The IRM is chosen over other masks such as the ideal binary mask since IRMs are shown to yield a better noise suppression performance (21).

This work investigates several DNN architectures which are trained to predict the IRM using biophysically-inspired representations of noisy speech. Essentially, the DNNs find an approximation of the complex function that maps the input features (i.e., the cochlear filtered acoustic input) to the IRMs. We make use of two popular DNN architectures: fully-connected and recurrent neural networks (RNN). The time-domain enhanced speech signal $\hat{s}[n]$ is obtained by multiplying the predicted IRMs with the noisy GT spectrogram (yields enhanced GT spectrogram $\hat{\mathbf{S}}$) followed by GT synthesis (16). The noise suppression quality of the enhanced signal $\hat{s}[n]$ with respect to the original clean speech signal $s[n]$ is then evaluated using the objective measures (i.e., PESQ, segSNR and CD).

We considered both fully-connected and the LSTM-based noise suppression systems in this study to investigate whether the improvements in speech enhancement in our approach result from enhanced input features to the DNN or the DNN itself. If biologically-inspired representations of noisy speech can improve both architectures, we have stronger evidence for attributing the improved noise suppression to the robustness of the speech representations themselves.

**Fully-connected NN-based approach.** Fully-connected DNN systems are comprised of multiple layers of matrix multiplications followed by a non-linear activation operation. Fully-connected speech enhancement systems predict the IRM of one T-F frame at a time. In order to provide some contextual information, a few neighboring frames are also fed to the DNN (known as frame expansion) and the IRM of the central frame is typically predicted. Such systems have been shown to outperform most of the conventional speech enhancement systems such as spectral subtraction and NMF (10, 22).

**RNN-based approach.** Although the temporal information in speech can be incorporated in fully-connected systems using frame expansion, these systems are not capable of explicitly modeling the relationship between the neighboring frames. RNNs, on the other hand, can capture the long-term contextual information (23) by means of recursive structures between the current and previous frames, and might consequently generate a better IRM estimate. We adopt RNNs that are based on long short-term memory (LSTM) cells (24) which act as a sequence-to-sequence model that generates the IRM sequences corresponding to the input feature sequences. LSTM-based systems are shown to yield a better speech enhancement performance over the fully-connected DNNs (11, 25).

**Evaluation metrics.** The speech enhancement performance is evaluated by comparing $\hat{s}[n]$ to the clean speech $s[n]$ embedded in the noisy speech stimulus using the following measures:

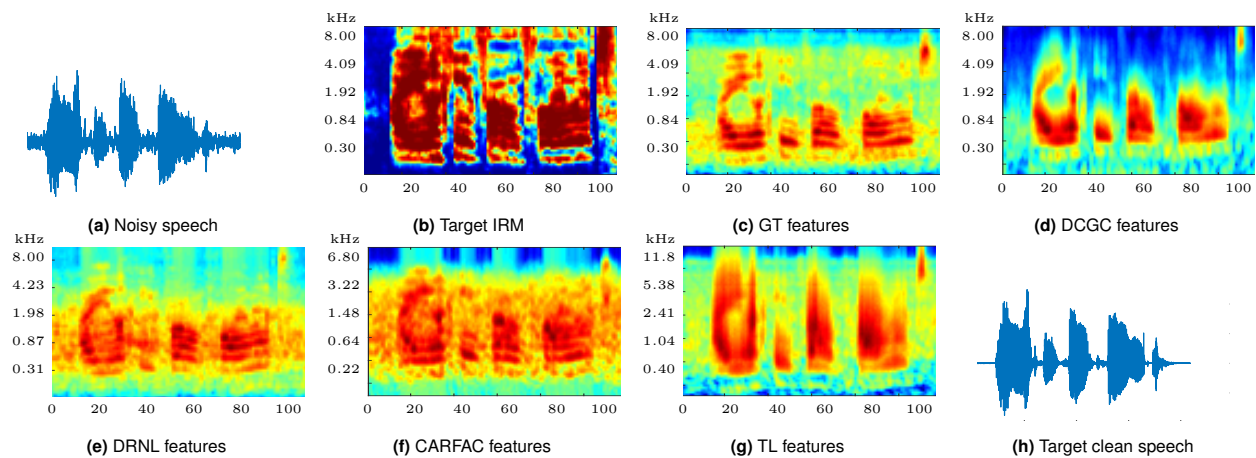

**Fig. 2.** Panels (a) and (h) depict an example noisy speech waveform and the corresponding target clean speech signal, respectively. In this example, babble noise was added at +3dB SNR. (b) The target IRM required to enhance the noisy GT spectrogram (computed using Equation Eq. (1) from the waveforms shown in (a) and (h)). (c) - (g) Input features obtained from filtering the noisy speech using the various cochlear models. The horizontal axis shows the frame indices on a linear scale whereas the vertical axis shows the center frequency of the filter-bank in kHz on a nonlinear scale depending on the frequency spacing used by the respective cochlear models.

perceptual evaluation of speech quality (PESQ) [37] in terms of mean opinion score (MOS), segmental SNR (segSNR) and cepstral distance (CD). The CD and segSNR measures are expressed in dB and were obtained using the implementations provided with the REVERB challenge (26). Higher values of PESQ and segSNR, and a lower value of CD indicate better performance.

For better readability, the improvements in these measures (shown as $\Delta$PESQ, $\Delta$segSNR and $\Delta$CD) when compared to the unprocessed noisy speech $y[n]$ are used for comparing the results. $\Delta$PESQ and $\Delta$segSNR are obtained by subtracting the metric obtained on the noisy speech (measured between $y[n]$ and $s[n]$) from that of the enhanced data (measured between $\hat{s}[n]$ and $s[n]$). The $\Delta$CD measure is obtained by subtracting the metric obtained for the enhanced data from that of the noisy data. In short, a higher $\Delta$ value implies a better performance for all the considered evaluation measures.

## Cochlear models

Five different cochlear models were placed between the noisy speech signal and DNN (see top row of the speech enhancement system in Fig. 1) to investigate whether incorporating cochlear speech processing improves the robustness of DNNs for noise suppression. The models vary in complexity and either form basic functional descriptions of cochlear filtering (e.g., GT filter-bank) or capture the cochlear mechanics and nonlinearities associated with human cochlear processing (e.g., transmission-line models). If biophysically-inspired models of cochlear processing yield more noise-robust input features to the DNN, it is expected that these models yield the best improvement in the objective evaluation metrics. We describe the main differences between the adopted models below, but refer the reader to (27) for further details on the cochlear model characteristics.

**Gammatone filter-bank (GT).** This model approximates human cochlear filtering using a parallel architecture comprised of bandpass filters with center frequencies between 50 and $f_s/2$ Hz, with $f_s$ the sampling frequency. The spacing between two center frequencies is given by the ERB-scale that describes the width of a single perceptually perceived auditory filter (19). GT filters can easily be inverted, which motivates their use for the enhancement phase of our processing (Fig. 1).

**Dynamically compressed Gammachirp (DCGC).** This model extends the GT filter-bank model by incorporating the nonlinear and compressive characteristics of cochlear processing associated with the dynamic range encoding of outer hair-cells in the cochlea (28). In a nutshell, the DCGC model consists of a set of parallel GT filters followed by a level-dependent high-pass asymmetric function which approximates the active and compressive action of cochlear outer hair-cells, which are responsible for stimulus level-dependent filter tuning.

**Dual resonance nonlinear filter-bank (DRNL).** This model builds on the GT filter-bank model by incorporating the outer-ear and middle-ear transfer functions (29). For each filter in the filter-bank, the DNRL model makes use of a dual resonance non-linear filter unit comprised of two parallel paths: a linear and a non-linear path. The outputs of these paths are summed up to simulate the nonlinear and level-dependent properties of cochlear filtering. In a nutshell, the DNRL model is comprised of a GT filter-bank, a compressive non-linearity and a second GT filter-bank.

**Cascade of asymmetric resonators with fast acting compression (CARFAC).** CARFAC is a cascade of second-order filters that are defined by one complex conjugate pair of zeros and poles (30). Whereas the previous models employed a parallel filter-bank architecture, CARFAC follows a serial (i.e., cascaded) approach in which each filter output serves as the input to the next (30). This architecture more realistically captures the mechanical structure of the cochlea in which sound travels from the base to the apex over the longitudinally coupled basilar-membrane (31). The individual filters act as second-order asymmetric resonators, whose characteristics (e.g., their center frequency and level-dependent damping ratio) are set to match human cochlear filter tuning.

**Nonlinear transmission line model (TL).** This model approximates cochlear processing as a cascade of shunt admittances

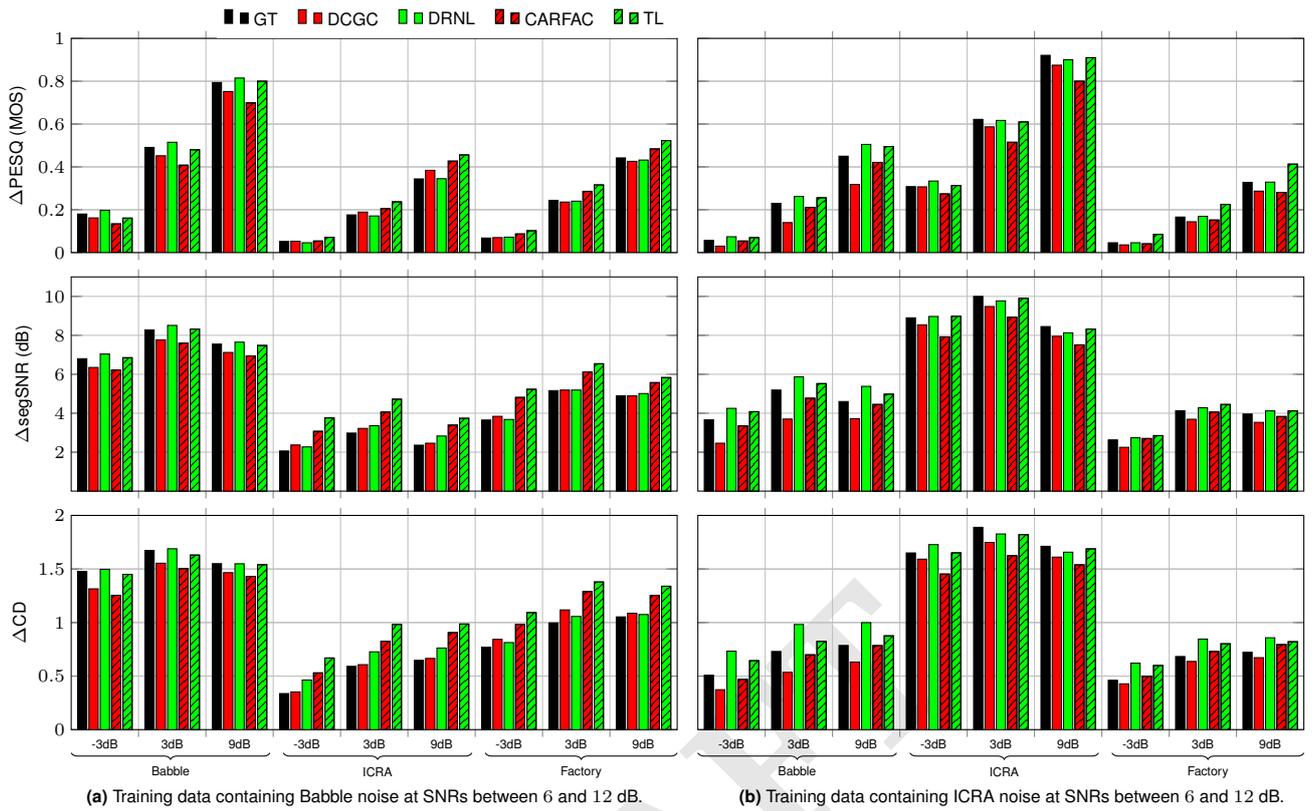

**(a)** Training data containing Babble noise at SNRs between 6 and 12 dB.  **(b)** Training data containing ICRA noise at SNRs between 6 and 12 dB.

**Fig. 3.** Comparison of noise suppression performance of the various fully-connected NN-based speech enhancement systems trained using the various cochlear models. The legends are same for all plots.

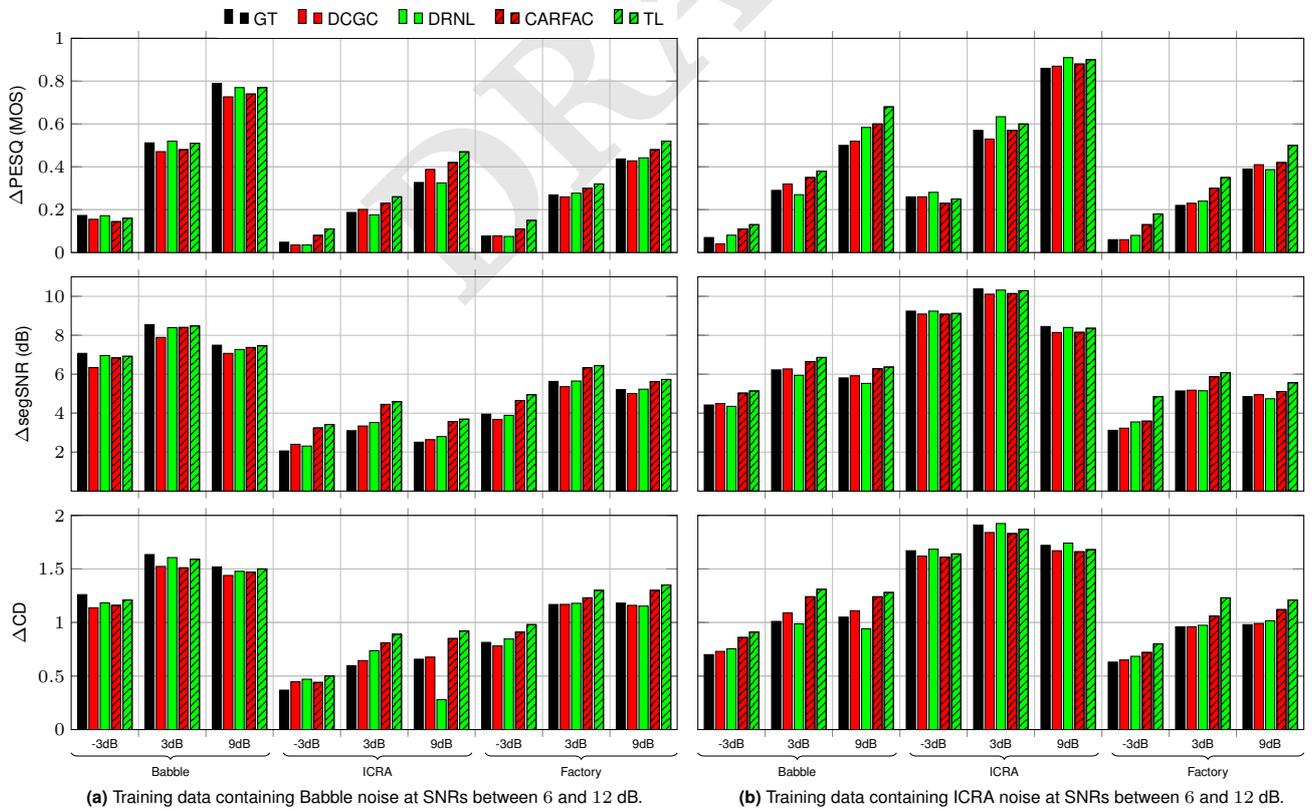

**(a)** Training data containing Babble noise at SNRs between 6 and 12 dB.  **(b)** Training data containing ICRA noise at SNRs between 6 and 12 dB.

**Fig. 4.** Comparison of noise suppression performance of the various LSTM-RNN-based speech enhancement systems trained using the various cochlear models. The legends are same for all plots.

and serial impedances that model the cochlear mechanical filter properties and fluid coupling, respectively. The model parameters are chosen to yield realistic human cochlear filter bandwidths that vary as a function of frequency and level (20, 32, 33). The cascaded organization of the bandpass filters results allows the modeling of several cochlear phenomena: e.g. two-tone suppression, frequency glides, and traveling waves which all result from the longitudinal coupling of the filters. The benefit of TL model over the other approaches is that they simulate forward and reverse traveling waves (i.e., otoacoustic emissions (34)). Human estimates of cochlear tuning, as derived from human otoacoustic emission measurements can thus be adopted to calibrate this model (33).

The cochlear-filtered noisy speech serves as input features to the DNNs in the noise suppression system. Figures 2 (c-g) show spectrograms of the input features derived from filtering the noisy speech signal (Babble noise added at 3dB SNR; Panel (a)) using the five considered cochlear models. Panel (b) shows the target IRM that was obtained from the noisy speech (a) and the clean speech signal (h) using Eq. (1). The aim of the DNNs is to find the best possible mapping (i.e., the mask $\mathbf{M}$) between the cochlear spectrograms (panels c-g) and the IRM (panel b).

## Results and discussion

To evaluate and compare the various speech enhancement systems under different training and mismatched testing conditions, we considered two training noise conditions and three noise test conditions with varying SNR levels.

The two training conditions were: 1) *TRAIN-BAB6to12dB* containing Babble noise added at an SNR between 6 and 12 dB, and 2) *TRAIN-ICRA6to12dB* with ICRA noise (35) added at an SNR between 6 and 12 dB. Thus, four different speech enhancement systems were trained for every cochlear model, i.e., two training sets each for fully-connected and LSTM-based noise suppression systems. Overall, this paper trained and compared 20 different speech enhancement systems.

The trained systems were evaluated for 9 noise conditions: Babble, ICRA and factory noises with $-3$, 3 and 9 dB SNR levels. Further details on the evaluation settings are provided in the Materials and Methods section. Since a training set contained only one noise type added at SNRs between 6 and 12 dB, the evaluation using 9 different noise conditions adequately validates the generalizability of the different cochlear models for mismatched noise and SNR conditions.

The noise suppression performance in terms of various speech quality measures on different test conditions are provided in Figures 3 and 4. The LSTM-based setting (Fig. 4) always outperformed the fully-connected setting (Fig. 3). This can be attributed to the ability of LSTMs in capturing long-term temporal dependence. In general, both the fully-connected and the LSTM-based speech enhancement systems show the same trends across different models and noise conditions. As expected, all models yielded a similar performance for matched noise conditions (i.e., those noise conditions that were present in the training data). The biophysically-inspired cascaded models (TL and CARFAC), yielded more generalizable speech enhancement systems as they consistently showed better performance in mismatched noise conditions over the more basic and parallel cochlear filter-bank models.

The DCGC model was found to outperform the GT model only in a few conditions and the improvements were not consistent across different training and test conditions. This can be attributed to the dynamic compression in the DCGC model which behaves differently for the different noise conditions. Additionally, Fig. 2 shows that the dynamic compression aggressively suppresses the high frequency regions. Lastly, the DRNL features were observed to overfit to the training data as the best performance was obtained for matched noise conditions whereas they failed to generalize to mismatched noise scenarios. Even though the DRNL is capable of strong noise suppression (eg., see Table I in (27)), a poorer performance for unseen noise conditions suggests that their suppression mechanism depends strongly on the noise type. Hence, including the nonlinear aspects of cochlear processing without capturing the cascaded architecture does not guarantee an improved speech enhancement.

In general, it can be seen that the cascaded models such as CARFAC and TL lead to better generalizable DNN systems when compared to the parallel filter-bank models. This shows the benefits and potential of using biophysically-inspired, cascaded filtering models of the cochlea for speech related applications. The reason why cascaded filter-models perform better than their parallel counterparts can be explained by the SNR improvement that is obtained when considering a single filter in the cascade. It was previously shown that the longitudinal coupling of filters results in a 2–5 dB SNR improvement at the filter output, for tone-in-noise stimuli (27). Our analysis shows that if the complexity of cochlear mechanics (yielding a natural noise-reduction) is adequately captured in the features provided to the DNN-based noise suppression systems, they become more generalizable and robust to different testing conditions.

Lastly, it is important to consider that advanced cochlear filter-models rely on nonlinear concepts (as in DNN architectures), but that these are part of a deterministic formulation. While describing the cochlear mechanics realistically require more computational effort than when using a parallel filter-bank, it comes with the benefit of more noise-robust speech enhancement systems.

## Materials and Methods

Clean speech recordings from the TIMIT dataset (16kHz sampling frequency) were used to simulate the various noisy speech scenarios. Three different noise conditions were considered: Babble, ICRA and Factory noise. ICRA is a non-stationary noise designed for clinical testing of hearing aids (35) with spectral and temporal characteristics similar to real-life speech and babble noise. The babble and factory noise recordings were taken from the NTT Ambient noise database.

From the 3696 utterances in the TIMIT training dataset, two training sets were generated, each containing Babble and ICRA noises that were added at a random SNR between 6 and 12 dB to the clean speech (referred to as TRAIN-BAB6to12dB and TRAIN-ICRA6to12dB, respectively). The test set was comprised of 9 noise conditions in which Babble, ICRA and Factory noises were added at $-3$, 3 and 9 dB SNR levels. The core test of the TIMIT database containing 192 recordings was used to create the test datasets.

The GT representation was extracted using the implementation provided in the auditory modeling toolbox (36). The DCGC and CARFAC representations were obtained using the auditory image modeling toolbox (AIM-MAT) (37) and the implementation provided in (38), respectively. The implementation for the TL model was obtained from (39). All these models were set to use $B = 64$ cochlear channels and the envelope energies were computed over a window-length of 20 ms shifted by 10 ms, resulting in 100 feature vectors per second. The logarithm of the energies together with their $\Delta$ coefficients were used as input to the DNN resulting in 128 features per frame. The training and test features were mean and variance normalized according to the mean and variance of the training set. The DNNs were trained to predict the IRMs corresponding to the 64 GT bands (ref. Fig. 1).

**Fully-connected NN setting.** The fully-connected DNN was comprised of 3 hidden layers with 1024 states per layer and sigmoid activation. A frame expansion of 3 left and 3 right frames were used yielding $7 \cdot 128 = 896$ dimensional input features. Drop-out technique with a keep probability of 0.9 was used to reduce over-fitting (40). The network was trained such that it minimizes the mean-square error between the output and the target IRMs using the Adam optimizer (41) for 20 epochs with a learning rate of 0.001 and batch size of 1024.

**LSTM-RNN setting.** This setting was comprised of 2 layers of LSTM-RNNs with 512 cells per layer and an output LSTM-layer containing 64 cells with sigmoid activation function to output the mask. This setting was trained using audio samples with a maximum duration of 5 seconds (500 frames) and the smaller recordings were zero-padded to match to the maximum duration. Recordings that were longer than 5 seconds were omitted from the training and test datasets, resulting in 3 577 training and 187 test utterances. Batch normalisation was used to obtain a faster convergence (42) and it was observed to yield a better noise suppression performance. The network was trained such that it minimized the mean-square error between the output and the target IRMs using the Adam optimizer (41) for 200 epochs with a learning rate of 0.0001 and a batch size of 16 recordings.

The dropout probabilities for both fully-connected and LSTM systems were chosen to yield the best results in the baseline system where GT features were used as the DNN input. For both approaches, a validation set of the same training noise type (but at 3 dB SNR) was used to store the model with the lowest loss on the validation set. The models were implemented using the Tensorflow toolkit (43) and NVIDIA Titan Xp GPUs were used for accelerating the training. The scripts for generating the noisy data from the TIMIT corpus and for training and testing the DNN systems are provided in the Github page https://github.com/HearingTechnology/BabyVerhulst_2018SpeechEnhancement. The users are required to have the TIMIT database with a license from the LDC (linguistic data consortium).

**ACKNOWLEDGMENTS.** This work was funded with support from the European Research Council under grant agreement No 678120 (RobSpear). We are also grateful to NVIDIA Corporation for donating the Titan Xp GPU used for this research.